# Modeling of the substrate influence on multielement THz detector operation

M. Sakhno, J. Gumenjuk-Sichevska, F. Sizov

Institute of Semiconductor Physics, National Academy of Sciences of Ukraine,

Nauki Av. 41, 03028 Kiev, Ukraine, e-mail: sakhno.m@gmail.com

**Abstract**

The development of THz multielement uncooled imagers based on focal plane arrays (FPAs) requires an optimization of the system parameters to achieve a homogeneous sensitivity of the array elements. Results of numerical simulation of the eight-element linear array of planar antennas with detecting elements, on a substrate of finite dimensions are presented. We establish how the substrate thickness $h$ and the relative permittivity $\varepsilon_r$ influence antenna pattern and antenna-detector matching for each element. We show that the antenna pattern depends on the detector position more than the antenna − detector impedance matching. The gain of array elements, the antenna-detector matching, and the homogeneity of the detector sensitivity can be simultaneously optimized by the proper choice of the substrate thickness $h$ and the relative permittivity $\varepsilon_r$. We show that multielement systems with large substrate thickness and high relative permittivity are not suitable for the imaging system implementation. To achieve uniform multielement system sensitivity, substrates with low permittivity ($\varepsilon_r <5$) and/or low thickness ($h< 60$ μm for the Si substrate) should be used. Finally, we investigate the operation of the detector array with optimally chosen substrate parameters together with the focusing lens, and show that the system is able to work as FPA without significant image corruption.

## 1. Introduction

In the field of terahertz (THz) technologies (ν ~ 0.1 – 10 THz), much attention is paid to the design of un-cooled efficient detectors, because of their potential importance for numerous applications in vision systems, spectroscopy, medicine, security, etc. Among the number of various detectors proposed and realized for THz/sub-THz spectral region in recent two decades (see, e.g., [1, 2]), there seem to be two most promising types of semiconductor detectors for the ambient temperature operation: Schottky barrier diodes (SBDs) and field effect transistors (FETs). They are relatively fast [3] and their operation rate in the THz range is limited only by the read-out electronics. Both of them have shown the capability of achieving noise equivalent temperature difference NETD~0.5 – 1 K [4], and were used as direct detectors and as nonlinear elements in heterodyne mixers [5, 6].

Silicon CMOS field-effect transistor FPA are well suited for multielement imaging system design, due to the possibility of integral implementation of a THz detection system with a matrix of ROICs. This design, based on the mature CMOS technology, makes possible to develop large-scale integrated THz devices similar to matrix imaging systems for visible and infrared ranges. FET transistors produced by the conventional technology cannot work in a usual mode (as an amplifier or a generator) at frequencies higher than their cutoff frequency, because their power gain is less than unity due to the parasitic effects and finite transit time. Nevertheless, for THz frequencies the signal rectification can occur in a narrow region of the transistor channel near the source, leading to a photoresponse of the transistor. FET transistor operation is usually described in terms of electron plasma waves in the transistor channel, after Dyakonov and Shur paper [7]. This direct detection operation has been first demonstrated experimentally for III/V FETs [8]. The applicability of CMOS FETs for direct THz detection was for the first time demonstrated in [9], in GaAs FETs [10]. The perspectives of silicon CMOS FETs as monolithic detectors integrated with antennas were shown in [11–13]. In papers [12, 14] usually states that thick substrates lead to substrate modes which absorb energy due to substrate conductivity. Thus antenna efficiency is lower on thick substrates. In this paper we show that different situation can appear for the case of thick finite substrate. Different elements can have different properties only due to different position on substrate even in the case of lossless substrate.

Despite the fact that a good quality image has been achieved by scanning the object in a focal plane by a single transistor [15], it still remains a challenge to obtain a good image by a large-format CMOS FPA with the optical system in the THz range. The new generation of multielement focal plane arrays (FPA) is expected to enable real-time imaging, reduce the scanning time and increase the information capacity and reliability of the video system by eliminating mechanical scanning components [1, 16]. Currently, there are no large-format THz-range FPAs with good spatial resolution capable of real-time operation. Chen et al [17] have demonstrated the ability to obtain an image in the transmission mode by 2×2 silicon-based Dicke radiometer FPA of the highest integration level, at 94 GHz signal frequency. [11], [18] have shown a possibility to scan an image at 591 GHz by the heterodyne detector 3-pixel line; however, this possibility remains virtual because in the two-dimensional case it is not possible to implement the differential detection scheme.

Homogeneity of the element characteristics in matrix or linear arrays is desirable. For the finite-size substrate with a high dielectric permittivity, the antenna gain may depend on the element position on the substrate. On the other hand, since the control buses on a chip are located differently with respect to different antennas, metallization of the buses can strongly affect the antennas operation [19, 20] and can lead to the gain and/or input impedance heterogeneity of the array elements. In [21] electromagnetic modeling has revealed the influence of length and orientation of bonding wires to the testing equipment on the spatial orientation of the two-lobe polarization pattern. Even if all pixels were identical, their radiation response and polarization dependence might be different due to different positioning on the substrate.

Typically, Si FETs or SBDs have dimensions about one hundred nanometers. Thus, in contrast to the visible and infrared range, where the direct detector element area is determined by the diffraction limit, in the THz range for an efficient input of the signal power into the detector one requires a receiving antenna with the area much larger than the detector itself. Usually, in the THz range, horns or planar antennas are used. Cat whisker antennas represented by very thin wires driven into semiconductor substrate were historically first used antennas with high-frequency SBDs back in the mid-sixties [22]. Horn antennas are used in sub-millimeter vision systems [23, 24] and custom detectors. They operate at frequencies up to 1.6 THz [25]. Horn antennas were used, e.g., in the William Hershel telescope [23]. Due to their high cost and complicated technology, they cannot be mass-produced. They have higher gain and lower losses than printed antennas and predictable characteristics due to mature technology, but they are bulky, expensive in manufacturing and poorly scalable for use in multielement detector systems. In contrast to that, planar antennas have worse characteristics but can be implemented by CMOS technology.

Printed antennas were first used in imaging system operated at 10 GHz [26]. Low efficiency of planar antennas is usually associated with substrate modes in infinite substrate case. Substrate modes and conductivity of the substrate contribute to energy losses and change of the antenna pattern [27–29]. A possible way to decrease losses due to substrate modes is to use high resistivity substrate or thin substrate [28].

Since the Maxwell equations are linear, already existing antenna designs (e.g. mm-wave) can be scaled to work in THz region. However, there are two aspects precluding such an approach: (i) the parameters of materials can change with the transition to a different frequency range; (ii) not all parameters can be scaled, for example due to technological limitations. Particularly, a substrate thickness in planar antennas cannot be scaled to a sufficient degree. Planar THz antenna is structurally similar to patch or printed dipole antennas of the radio frequency range [27]. However, the existence of electrically thick (the substrate thickness $h > 0.1 \cdot \lambda_d$, where $\lambda_d$ is the wave length inside the dielectric substrate) finite size substrate may be crucial for antenna operation in the THz frequency range.

It is clear that there are many other parameters which may affect the antenna operation: length, width, and thickness of the substrate, the antenna position, cross talk of different antennas, and the lead contacts influence. In fact, every antenna operation depends on many variables. Finding the impact of all parameters together is too complex to be analytically tractable. It is generally difficult to separately control the influence of every parameter and to ensure the stability of the optimal solution. A small change in one of these parameters may lead to a strong change in the characteristics of the entire system. In this work, we check and prove the assumption that the substrate thickness and relative permittivity are the critical parameters for the technology which have the most significant impact on the operation of the detector array. We establish the range of those two parameters where the behaviour of the antenna becomes predictable and stable. This is consistent with the previous work for the lower frequency range [27], showing that modeling the antenna using the approximation of infinite (in plane) substrate is often appropriate for electrically thin substrates.

Historically, the noise equivalent power (*NEP*) is used for detector characterization. It is a useful figure of merit in visible and infrared region. There is a subtlety in using *NEP* for characterization of detectors in THz range: to determine the NEP, one should know the effective area from which the power is introduced to the detector. Traditionally, this area is determined using the pitch size of the detector ($S^{'}$) [30, 31]. The measured *NEP$_{meas}$* includes the electrical *NEP$_{el}$* of detector itself, the antenna gain $G$, and matching between the antenna and the detector $\eta_A$:

$$NEP_{meas} = NEP_{el} \frac{S^{'}}{G \frac{\lambda^2}{4\pi} \eta_A} \tag{1}$$

This formula assumes illumination by the plane wave. If a focusing lens is used, then antenna and lens should be considered as a whole, and the gain should be determined for the whole system. Different parts of this formula (1) are discussed in different sections: gain G in section 3, antenna-detector matching $\eta_A$ in section 4, operation with lens in section 5. Antenna parameters cannot be measured independently [32]. The dependence of *NEP$_{meas}$* on the pitch size presents a difficulty, since different authors use different definitions of the pitch size [12, 33] which may result in values varying by up to 2 orders of magnitude. Thus, *NEP$_{meas}$* becomes strongly dependent on the method of calculating the pitch size and not a characteristic of detection element itself. It should be noted that unlike the radio frequency range, in the THz range antenna measurements are usually performed in receiving mode because of absence of THz generators which could be integrated with the antenna instead of a detecting element [32]. In this case, an antenna and a detector are measured as a whole. It is very difficult or impossible to extract parameters of antenna and detector from the measurements independently.

One can see from (1) that to improve NEP of the system, the antenna should have maximal gain and the detector impedance should be matched with the impedance of the antenna [1, 17]. The only way of analyzing the influence of different parameters in Eq.(1) in the case of planar antennas on the NEP is the numerical simulation, because of the sheer complexity of the system.

The objectives of this paper are (1) to check if it is feasible to reach a homogeneous sensitivity in the multielement detecting array on a substrate of finite dimensions by a proper choice of the substrate thickness $h$ and the relative permittivity $\varepsilon_r$; (2) to make sure that the array with such suitably chosen substrate parameters will work as FPA together with the focusing lens without image distortion.

We performed numerical simulation for the linear array of eight elements with modified bow-tie antennas on a finite-size substrate, the design of which is described in Section 2. In Section 3 we examine how the antenna position on a finite substrate influences the detector parameters, and compare antenna gains $G$ of different elements for the cases of electrically thick and thin substrates. We prove that it is possible to optimize the gain of the array by the proper choice of substrate thickness and relative permittivity. In Section 4 we find out how the antenna – detector matching coefficient $\eta_A$ for different array elements depends on the substrate properties, and how it influences the *NEP*. In Section 5 we study a vision system, based on the considered multielement array of silicon FETs (with properly chosen substrate parameters) integrated with focusing lens, and show that it is able to operate in FPA or scanning mode.

## 2. Simulated system description

The system considered here is a simplified model of FET FPA which could be implemented by CMOS technology. The simulated system is a linear array of 8 detector elements with identical antennas on a finite-size dielectric substrate. The elements are located symmetrically relative to the substrate center (Fig. 1). The substrate backside is coated by metal. The substrate width is 1 mm, length is 10 mm. The distance between the element centers is 1 mm which is roughly equal to the diffraction limit at 300 GHz. Several substrate thicknesses were modeled: 50, 140, and 650 μm: 50 μm is the minimal reasonable substrate thickness in CMOS technology (thinner substrate is fragile), 140 μm can be obtained by back thin, and 650 μm is one of the typical thickness for the standard CMOS technology. Dimensions of a transistor are much smaller than the radiation wavelength, thus it is considered as a lumped element (see Sec. 4 for the effective equivalent circuit scheme). The system parameters were intentionally idealized to make clear that observed effects are due to antenna position and not due to the skin-effect, dielectric losses in substrate, etc. That is why we neglected electric losses in the substrate, so its conductivity was assumed to be $\sigma=0$. The model is highly simplified, but we believe it is able to capture the complexity of the implementation of linear or matrix arrays layout on substrates with high dielectric permittivity like silicon ones.

Properties of the system with two printed antennas on *infinite* substrate plane were considered in [29] and it was shown that the two antennas have equal characteristics. Effect of substrate size on the parameters of a single antenna was modeled in [27], and it was shown that the thinner substrate is a better option. In our case, when the substrate has finite width and length in plane, the geometrical equivalence of antennas does not lead to the uniformity of their characteristics because of the difference in their arrangement relative to substrate borders.

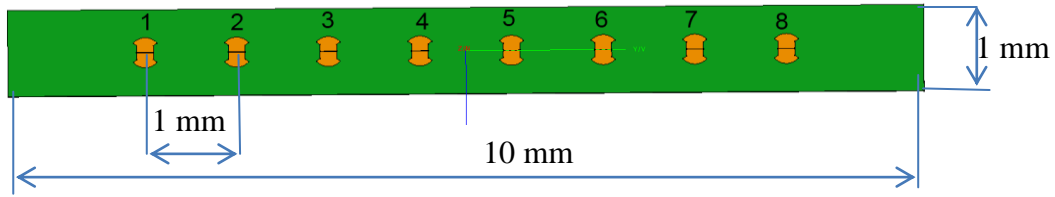

**Fig. 1** The modeled system design: 8 antennas on a substrate of finite size. Antennas are positioned symmetrically relative to the substrate center

Each element is a modified bow-tie antenna. The free-space classical bow-tie antenna is frequency independent [27] and its properties are determined solely by its angles, but in our case the system design includes rectangular contact areas on the antenna and finite size substrate as shown in **Fig. 2**. A side effect of this is that the antenna becomes frequency dependent. Moreover the antenna on a substrate can be frequency independent only if the substrate electric thickness $\sqrt{\varepsilon} \cdot h$ is much smaller than the signal wavelength [27]. Parameters of a single element detector were optimized earlier for maximal gain at 300 GHz in [34] for planar infinite substrate with the thickness of 500 μm. As it was shown, the signal interference inside the substrate leads to the drastic narrowing of the operating frequency range of the system, when the antenna is placed on a substrate with the thickness of the order of the wavelength (for the standard CMOS technology the substrate thickness is about 400-650 μm). Parameters for maximum gain in the normal direction of the antenna at 300 GHz are: $a_1 = 10 \mu m$, $a_2 = 75.8 \mu m$, $r = 164 \mu m$, $d = 20 \mu m$, $\phi = 104°$ (see **Fig. 2**).

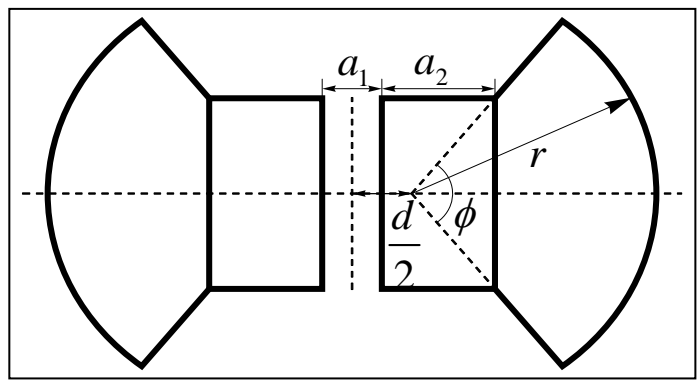

**Fig. 2** A sketch of a single bow-tie antenna used in the present model

Since in the THz range the substrate thickness is comparable with the radiation wavelength, the transmission line or cavity models [27], which are appropriate for the substrate thickness $\sqrt{\varepsilon}h = (0.001 - 0.01)\lambda$, become inapplicable. Therefore, modeling such antennas on the substrate requires using full-wave methods, such as MOM (Method of Moments), FEM (finite element method) or FDTD (finite-difference time-domain) method. The commercial program EMSS FEKO [35] was used for the modeling due to the availability of MLFMM (the Multilevel Fast Multipole Method specially created for electrically large structures).

The antenna array was modeled only for the first four elements, due to the mirror symmetry of the system. Each antenna was modeled in the transmitting mode. To analyze the behavior of a particular antenna in this multi-antenna system, other antennas were loaded with the impedance of 50 Ω. Matching the antenna and transistor impedances is discussed in Sect. 4.

## 3. Antenna pattern: the dependence on the substrate permittivity

In this section the influence of the substrate thickness and permittivity on the antenna pattern is discussed. To maximize the vision system response [36], the antenna should have (i) a single main lobe normal to the surface (this is desired to ensure the correct displaying of the focal plane points); (ii) as weak side lobes as possible; (iii) antenna diagrams and impedances should be identical for different antennas of the array. Further, the form of the lobe should be close to the Gaussian for better optical system matching. Substrates with different thicknesses and permittivities were checked in this work for the compliance with these conditions.

Dipole antenna operation on electrically thick substrate has been studied in [37]. For the finite size substrate, gain dependence on the substrate width and length should be observed.

Figs. 3-5 present the results of the modeling of the linear gain for every element of the array for substrate thicknesses of *50 μm*, *140 μm* and *650 μm*, respectively. For each value of the thickness, the substrate permittivity $\varepsilon_r$ was changed from 2 to 12 with the step of 1, and gain diagrams for each of the first 4 antennas were obtained. For the remaining 4 elements of the array the picture is mirror symmetric, so they are not shown in the figure for the sake of clarity. Here we consider the linear gain of antenna, because it is more sensitive to the parameters change then the logarithmic one (dBi). Patterns for different antennas were positioned on one image to facilitate the comparison of different elements. The following peculiarities were observed: When the substrate thickness is small (*h=50 μm*), antenna diagrams for different elements of array are very similar (**Fig. 3**) and there are no side lobes for all 3 values of the substrate permittivity. Patterns for different antennas become different at the substrate thickness *140 μm* (Fig. 4), and side lobes appear with the increase of the permittivity. Degradation of antenna pattern is weak for *h=50 μm* (**Fig. 3**) and becomes more pronounced for *h=140μm* (Fig. 4).

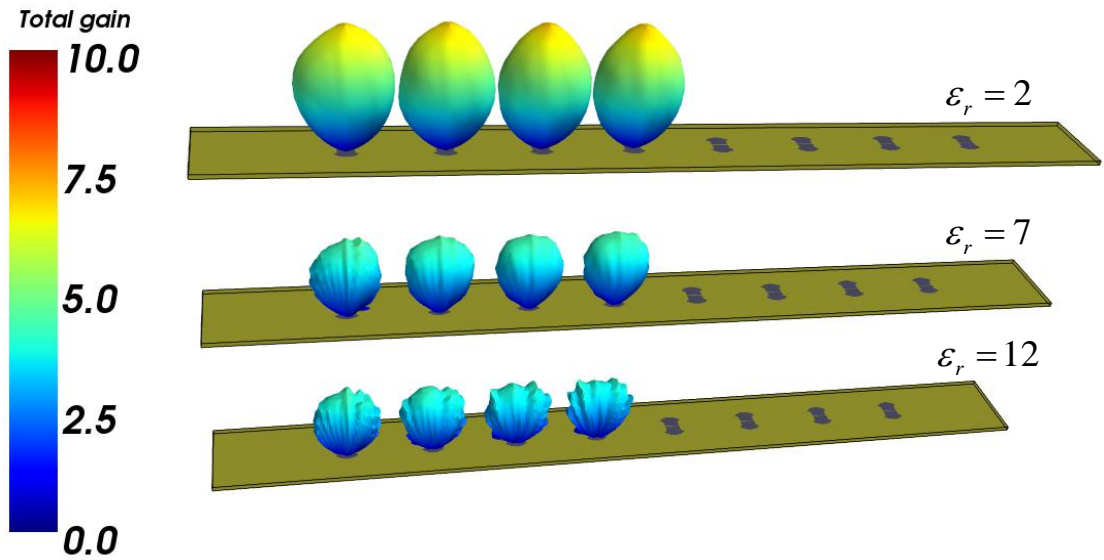

**Fig. 3** Linear gain diagram for the radiation frequency 300 GHz. Only the first four elements of the linear array are shown. Substrate thickness is $h=50$ μm, three values of the permittivity $\varepsilon_r=$ 2, 7, 12 are shown. Each antenna was simulated and the results were combined on one picture to facilitate the comparison of different elements

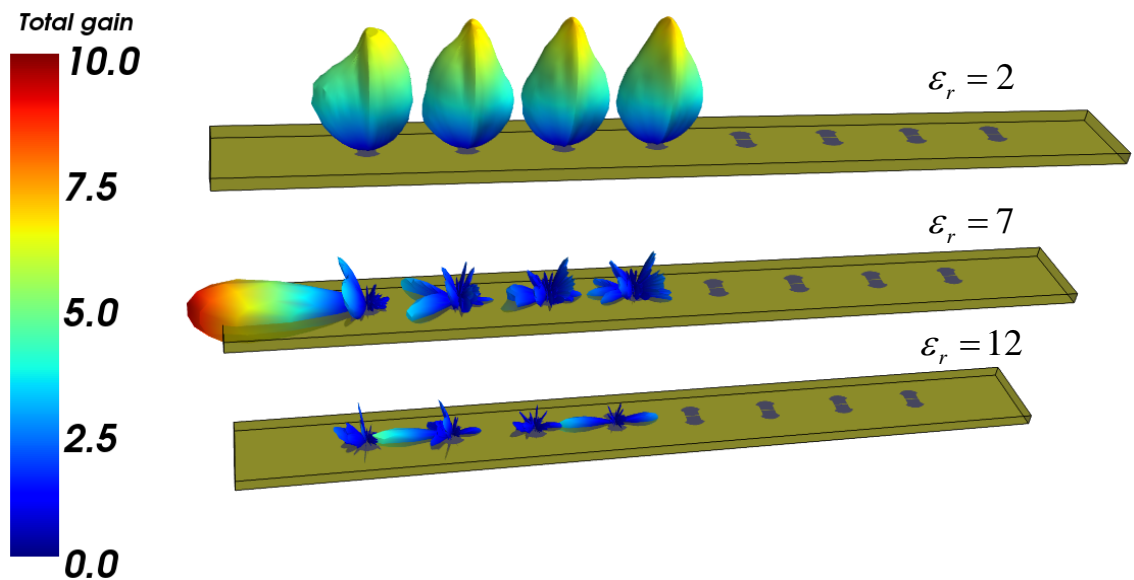

**Fig. 4** The same as in Fig. 3, but for the substrate thickness $h=140$ μm. Degradation of antenna pattern is observed for the substrates with the permittivity $\varepsilon_r>5$

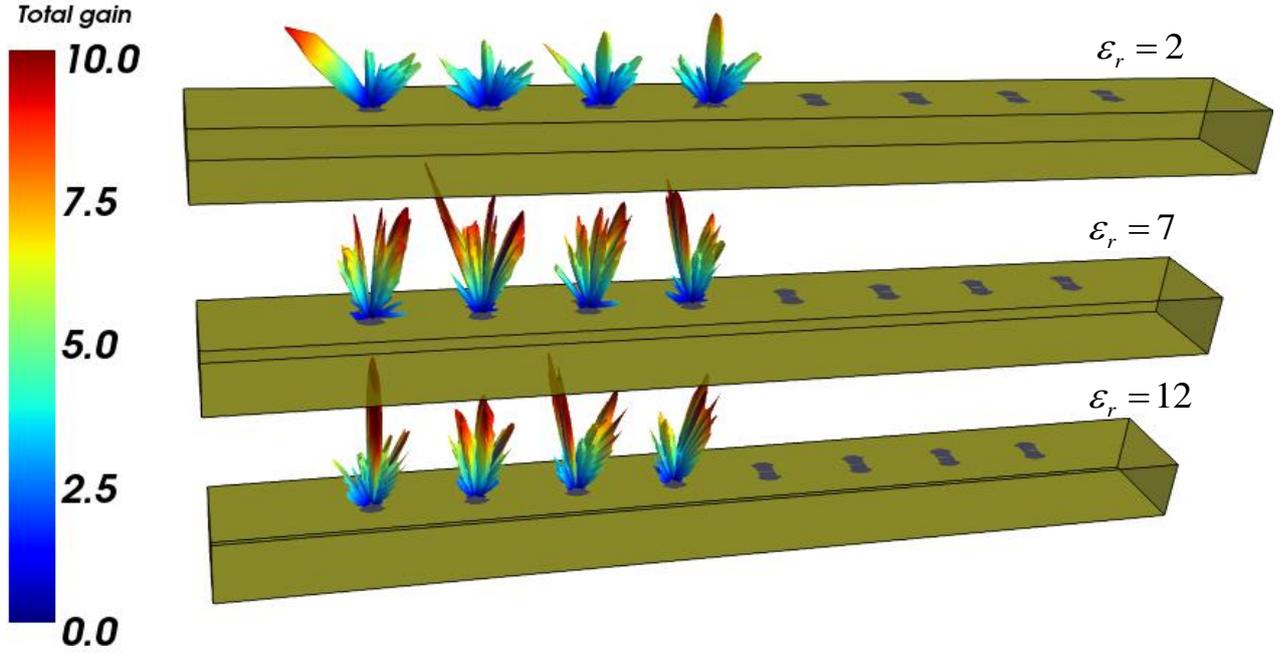

**Fig. 5** The same as in Figs. 3 and 4, but for the substrate thickness $h= 650\mu m$. Degradation of antenna pattern is observed even for low permittivity ($\varepsilon_r=2$) of the substrate.

Substrate thickness is a key parameter, which determines importance of other parameters. When electrical thickness of the substrate is low than there is no dependence of antenna patterns on antenna position. This supports usage of infinite substrate approximation for thin substrates. For thick substrates dependence on antenna position or "edge effects" becomes important. For $h=650$ $\mu m$, (see **Fig. 5**) degradation of antenna pattern is observed even for low permittivity ($\varepsilon_r=2$) of the substrate and becomes worse with the increase of $\varepsilon_r$. The maximal gain lobe is oblique to the surface and directed differently for different element.

To explain this phenomenon, we analyse the presence of substrate modes by determining the cut-off frequencies of modes in infinite substrate for different substrate thicknesses and permittivities. Cut-off frequencies of infinite substrate modes are given by [29]:

$$f_{cn} = \frac{nc}{4h\sqrt{\varepsilon_r - 1}}, \tag{2}$$

where $n$ is the mode number, $h$ is the substrate thickness, $\varepsilon_r$ is the relative permittivity of the substrate, and $c$ is the speed of light in free space. The $TM_0$ mode with zero cut-off frequency is always present in the infinite substrate. However, the energy of this mode is lowered with decreasing $h$, so for thin substrates with $h < 0.01\lambda_0\varepsilon_r^{-1/2}$ the coupling with substrate modes is not important [29]. The presence of substrate modes leads to degradation of antenna pattern in the case of infinite substrate [29]. In the case of finite-size substrate, the determination of substrate modes is much more complicated [28], so we determine infinite substrate modes.

From Eq. (2), the cut-off frequency for the first mode $f_{c1}$ can be estimated (see Table 1). At $h=50\mu m$ for all considered permittivies $f_{c1}>300GHz$, thus only the zero mode is exited. Substrate modes are not important and do not produce antenna pattern changes (as seen in **Fig. 3**) for such a thin substrate. At $h=140\mu m$ for $\varepsilon_r=2$ only the zero mode is excited, but for $\varepsilon_r=7$ and $\varepsilon_r=12$ other modes can be excited for the signal frequency 300 GHz, and antenna pattern degradation arises (as seen in **Fig. 4**). For $h=650\mu m$ and $\varepsilon_r=2, 7, 12$ $f_{c1}<300GHz$. This leads to antenna pattern degradation (as seen in **Fig. 5**) for all permittivities of the substrate. Depending on the antenna position on the substrate, the antenna interacts differently with the substrate modes. A larger number of presented modes in the substrate leads to more complicated interaction between the modes and the antenna. For the electrically thin substrate the size in plane does not matter. For the thick substrate every dimension of it influences on antenna pattern. Thus it is necessary to deal with electrically thin substrate to make optimization possible. In this case we need to optimize only thickness of substrate with the corresponding $\varepsilon_r$.

Table 1. Cut-off frequency of the first mode $f_{c1}$ for modelled substrate thicknesses and permittivities for infinite substrate

| h, μm | $\varepsilon_r=2$ | $\varepsilon_r=7$ | $\varepsilon_r=12$ |
|---|---|---|---|
| 50 | 1.5THz | 0.612 THz | 0.452 THz |
| 140 | 0.536 THz | 0.219 THz | 0.162 THz |
| 650 | 0.116 THz | 0.047 THz | 0.035 THz |

The results shown in Figs. 3-5 are summarized in Fig. 6. For thin substrates and low $\varepsilon_r$, the normal gain is high and is almost the same for different elements. With increasing $\varepsilon_r$, the normal gain decreases, the difference between elements appears (**Fig. 6**), and at the same time multiple side lobes emerge in the antenna diagram (**Fig. 4**). The decrease of the normal direction gain with the increase of $\varepsilon_r$ was also observed for antennas on infinite substrate in [38]. It was explained by the increase of the number of substrate modes and, consequently, of the energy loss through them. The frequency dependence of the total gain in the normal direction of the 1[st] and 4[th] array elements is shown in **Fig. 7**. The frequency dependence for different elements is different for high permittivity substrate ($\varepsilon_r = 12$), which is undesirable, while for $\varepsilon_r=2$ the frequency dependence is practically flat and the gain is much higher than for substrate with $\varepsilon_r=12$.

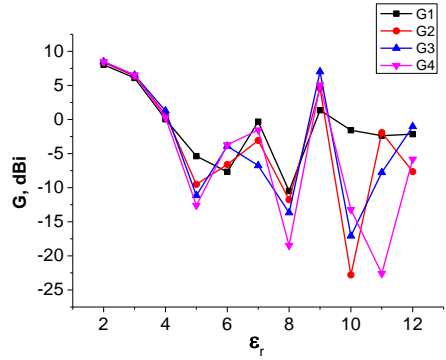

**Fig. 6** The dependence of the calculated total antenna gain $G$ in the normal direction on the substrate permittivity $\varepsilon_r$ for four different array elements (labelled $G$1 to $G$4). The substrate thickness is 140 μm and the radiation frequency is 300 GHz. When the permittivity is small ($\varepsilon_r \leq 4$) the difference between antennas is small and the gain is high

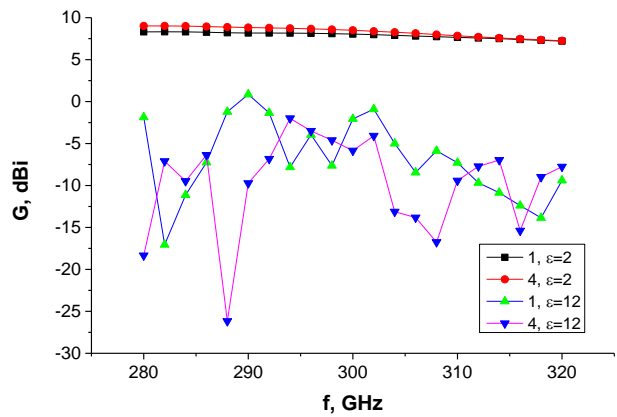

**Fig. 7** Calculated normal gain frequency dependence for 1st and 4th elements on 140 μm thick substrate, for different values of substrate permittivity: $\varepsilon_r =2$ and $\varepsilon_r =12$

To summarize the modeling results, one can observe that with the increase of the substrate permittivity $\varepsilon_r$

1) the gain $G$ in the normal direction decreases and its main lobe becomes inclined to the substrate, multiple side lobes appear (Fig. 4, Fig. **5**);
2) the difference between antennas appears (Fig. 4, Fig. 7);
3) the gain becomes frequency dependent (Fig. 7).

Thereby, the multielement system with high substrate electric thickness and high $\varepsilon$ is not suitable for the imaging system implementation. In addition, care should be taken when comparing parameters of elements created on large substrates. In general, it is shown that the presence of the substrate modes leads not only to energy dissipation in a thick substrate, but to the degradation of the antenna pattern.

It should be remarked that realistic CMOS systems are more complicated than the one we have modeled. A substrate usually has some conductivity (doping dependent). There are a lot of metallic elements on a substrate which can interact with antennas. It also should be dependent on experimental sample setup (chip package, external contacts, etc.). So, rigorous modeling demands high computational resources. The key result of our modeling is that the electrically thin substrate is necessary for the detector implementation. Our simulation results are supported by the experimental data of [5, 39] who have used very thin dielectric substrate layers with interconnecting metallization. Alternatively, one can design arrays on their own separate small dielectric substrates and use hybrid technology for implementation FPA on common substrate with low permittivity [40, 41].

## 4. Antenna–detector matching

In this section, we discuss matching of the antenna input impedance with FET, which is important for the optimization of the detector system.

The concept of FET THz detection [30] is based on the excitation of plasma waves in the two-dimensional electron gas in the transistor channel, which plays the role of a nonlinear element. When the THz radiation power is introduced to FET (e.g., between the gate and the source), the THz signal AC voltage modulates simultaneously the carrier density and the carrier drift velocity in the channel. In this sense, the detection mechanism at THz-range frequencies is similar to the signal detection at low frequency when the signal is proportional to the input power (i.e., to the square of the voltage) [42]. The signal detection is due to FET nonlinearity that leads to rectification of the AC induced by the THz radiation. Thus, a non-resonant DC voltage response $\delta V_{DS}$ between drain and source arises that is proportional to the THz power received. The FET design is usually symmetrical (source and drain regions are identical). For the appearance of the DC $\delta V_{DS}$ voltage, the asymmetry between the source and drain is required. The transistor cannot operate as an amplifier or generator at THz frequencies since its power gain is less than unity due to the parasitic resistance. Nevertheless, the signal with a much lower modulation frequency could be well detected and rectified in a short region near the source, since it is not affected by parasitic effects. Discussion of detection mechanism can be found elsewhere [43–45]. Here we are interested in the input impedance of FET only.

Dimensions of the field-effect transistor are much smaller than the wavelength in THz or sub-THz, so FET can be described as lumped element. In the case of ideal antenna and detector impedance matching ($Z_{tr} = Z_{ant}^*$, where $Z_{ant}$ is the antenna input impedance and $Z_{tr}$ is the input impedance of the transistor between source and gate), the maximum power on the load (detector) connected to antenna illuminated by the plane wave is [27]:

$$P_{\max} = \frac{G\lambda^2}{4\pi} I, \qquad (3),$$

Where $G$ is the antenna gain, $I$ is the radiation intensity, and $\lambda$ is the wavelength. The power released on the detector is proportional to the squared antenna voltage

$$P_{max} = \frac{|U_{ant}|^2}{8\,\mathrm{Re}\,Z_{ant}} \tag{4}$$

From (3) and (4), one obtains the voltage on antenna terminals as

$$|U_{ant}|^2 = \frac{2}{\pi} \cdot G\lambda^2 I \cdot \mathrm{Re}\,Z_{ant} \tag{5}$$

In the case of impedance mismatch, when $Z_{ant} \neq Z_{tr}^*$, the power on the load is

$$P = \frac{G\lambda^2\,\mathrm{Re}\,Z_{ant}\,\mathrm{Re}\,Z_{tr}}{\pi|Z_{tr}+Z_{ant}|^2} I \tag{6}$$

Thus, the impedance matching coefficient, which is the ratio of the power released on detector to the maximum power that can be released, is obtained as follows

$$\eta_a = \frac{P}{P_{max}} = 4\frac{\mathrm{Re}\,Z_{ant}\,\mathrm{Re}\,Z_{tr}}{|Z_{tr}+Z_{ant}|^2} \tag{7}$$

The optical responsivity $R_V^{opt}$ (the ratio between the signal from the detector and the maximum power that can be obtained from the antenna) is connected to the electrical responsivity (the ratio between the detector signal and the power released in the detector) $R_V^{el}$ which is determined for FET in [45] and for SBD in [46]

$$R_V^{opt} = R_V^{el}\eta_a \tag{8}.$$

The electrical equivalent circuit of antenna + FET system is presented in **Fig. 8**. FET can be separated into two parts [47, 48]: (*i*) an intrinsic part (where the signal rectification occurs) with the impedance $Z_{SG,int}$, and (*ii*) an extrinsic one, related to the parasitic capacitance $C_P$, the parasitic gate resistance $R_G$ and the source series resistance $R_S$. The transistor input impedance, resulting from the parallel connection of these two parts, is

$$Z_{tr} = R_S + R_G + \frac{1}{j\omega C_p} \Big\| Z_{SG,int} \tag{9}.$$

The resistances depend on transistor dimensions as follows: $R_G \sim W/L$, $R_S \sim 1/W$, where $L$ and $W$ are the transistor channel length and width, respectively, the parasitic capacitance $C_P \sim W$, and

$j = \sqrt{-1}$. For the internal FET impedance we use the result obtained from the transmission line channel model [49]

$$Z_{SG,int} = \sqrt{\frac{R_{CH}}{j\omega C_{ox}}}, \qquad (10)$$

where $R_{CH}$ is the FET channel resistance and $C_{ox}$ is the under-gate oxide capacitance.

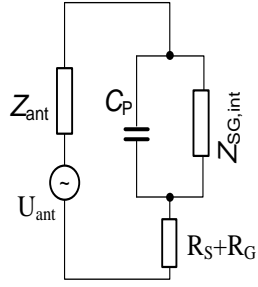

**Fig. 8** Electrical equivalent circuit of the antenna with FET as a detector. $Z_{ant}$ is the antenna impedance; $U_{ant}$ is the radiation voltage amplitude at the antenna; $R_S + R_G$ is the parasitic resistance of FET, $C_P$ is the parasitic capacitance; $Z_{SG,int}$ is the internal source-gate impedance

The active part of FET input impedance is dependent on $W$, $L$ and the technological design rules. In the case of the high frequency $\omega \gg \frac{C_{ox}}{R_{CH} C_P^2} \approx 2\cdot 10^9 s^{-1}$ (see Eq. 8), the internal input impedance $Z_{SG,int}$ is much larger than the impedance of the parasitic capacitance $jX_P = \frac{1}{j\omega C_P}$, and $|Z_{SG,int}| \gg |X_P|$, so the transistor impedance is mainly determined by parasitic effects

$$Z_{tr} = R_S + R_G + jX_P \qquad (11)$$

Using the above formulas, we have calculated the impedance matching coefficient for 1-μm technology FET with the channel dimensions $W/L = 20/2$ (μm) and antennas considered in Sec. 2. The FET parameters were taken from BSIM3 model files and technology data as follows: $R_G = 150$ Ω, $R_S = 50$ Ω, $C_p = 4$ fF. At the radiation frequency 300 GHz, Eq. (9) yields the FET input impedance $Z_{tr} = (200 - j\cdot 130)$ Ω. The impedance matching coefficient $\eta_a$ calculated according to Eq. (5) is shown in **Fig. 9**. It stays approximately the same for different elements, even for electrically thick substrate ($h = 650$ μm, $\varepsilon_r = 12$). It is possible to maximize the impedance matching by varying the transistor dimensions or by use of the transmission line. Thus, the gain dependence on the detector position on the finite substrate is the main factor determining the variation of the measured *NEP* (see Eq. (1)). Almost 100% matching is obtained for $h = 50$ μm and $\varepsilon_r = 4$ at $f = 300$ GHz (see the left panel of Fig.9). Substrate parameters for the optimal antenna-detector matching and for antenna homogeneity are different. Electrically thicker substrate may have better matching with the antenna depending on transistor dimensions (please, compare matching for $h = 50$ μm, $\varepsilon_r = 2$ and 4 and for $\varepsilon_r = 2$ $h = 50$ μm and 140 μm). Optimal substrate thickness should be a compromise between elements homogeneity and matching with antenna.

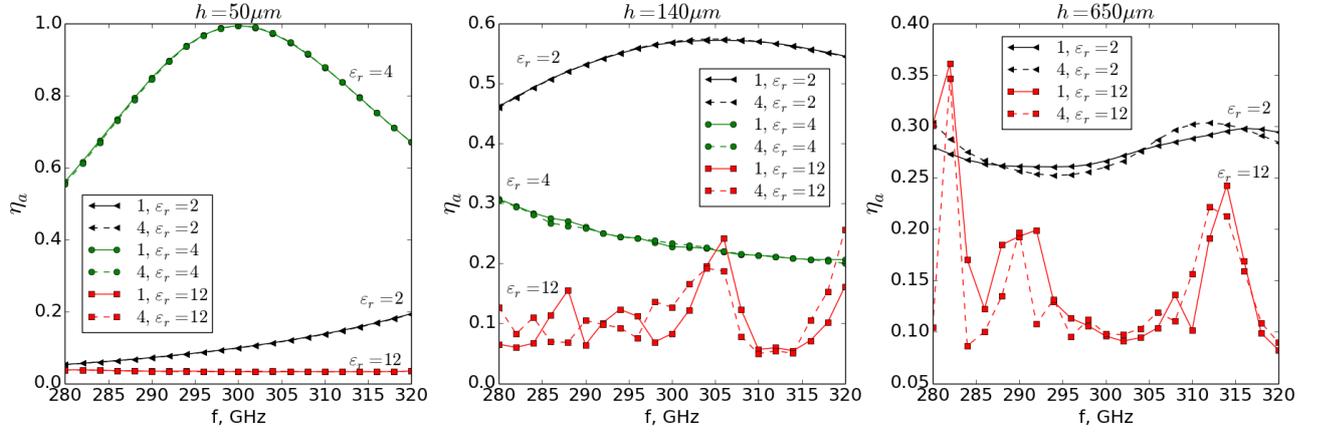

**Fig. 9** The impedance matching coefficient for 1$^{st}$ and 4$^{th}$ array elements versus the radiation frequency, for different substrate thickness. In the left panel, curves coincide for h=50μm.

For SBD with an antenna, the matching condition is also determined by Eq.(7) with $Z_{tr}$ substituted for $Z_d = R_S + \dfrac{1}{j\omega C_p} \| R_d$ [46], where $R_s$ should be interpreted as the series diode resistance, $C_p$ as the diode junction capacity, and $R_d$ as the junction resistance. For a typical zero biased SBD $Rs \approx 10$ Ω and $C_p \sim 5\ldots 10$ fF, $R_d \sim 3$ kΩ. Thus, similarly to FETs, for a typical SBD in a linear array implementation, the impedance matching coefficient has only a weak dependence on the element position (for typical values of detector input impedance as for any type of detectors with antennas with similar input impedance), and the gain dependence on the detector position on the finite substrate is the main factor of different sensitivity of detectors in array.

## 5. Imaging system with a lens

Focal plane array operation principle is that every object element screens by the FPA element. Each receiving element of the array is responsible for an image section and thus represents a pixel of the image, so each antenna of FPA, together with the optical system, should have its own narrow direction of maximal gain. Usually the interaction of the antenna and optics is described in terms of quasioptics (Gaussian beam propagation) [50] and was extensively studied in e.g. radio telescope design. Here we use another approach modeling the entire system. The main reason is that the lens is placed in the antenna near field region, the edge of which is determined by the radius $R_2 = 2\dfrac{D^2}{\lambda} \approx 20 cm$ (where D=10 mm is the length of linear array, λ=1 mm) [27] and, therefore, the quasi-optical formalism cannot be applied. D is maximum dimension of antenna. Antenna dimensions is determined by region where non-negligible current occurs. We use substrate length as this distance. $R_2$ – is radius of beginning far-field region of the antenna. Geometrical optics uses assumptions of plane wave so lens should be in far-field zone of the antenna. We use distance $R_2$ to check applicability of the geometrical optics.

In Sec. 2 it was assumed that the system is illuminated by a plane wave. In contrast to that, here we consider the situation when the detector array is placed in the focal plane of the lens and each detector of FPA is illuminated by a spherical wave. The aim of this simulation was to prove that the system (on a thin substrate) can operate as FPA, as well as in devices with mechanical scanning.

An antenna in a mechanical scanning system, when being moved in the lens focal plane, should give the signal proportional to the illumination intensity at each point of the plane. It is not obvious because the effective size of antenna can exceed the electrical field distribution length in the Airy disk. The resulting signal in the detector is a convolution of the field distribution function in the lens focal plane and the detector pattern. In the optical range, the detector pattern is simple due to the detector sensitivity independence of the angle of light incidence. This is no more the case in the THz range because of the presence of the receiving antenna whose gain is angle-dependent.

The Plano Convex lens with relative permittivity $\varepsilon_{r,lens} = 4$ and a focal ratio $f/\#=0.625$ was modeled. Here the lens diameter $D_l$ is 16 mm, the lens thickness is 4 mm, the front lens surface curvature radius $r_l$ is 10 mm and the focal length $f_l$ is 10 mm. The chosen dimensions of the lens are a compromise between computational resources needed for the modeling and the lens gain. This lens configuration is similar to the experimental setup of [51] **Fig. 10**a. The aim of our modeling was to show how the lens interacts with the antenna array and whether the system will operate as FPA.

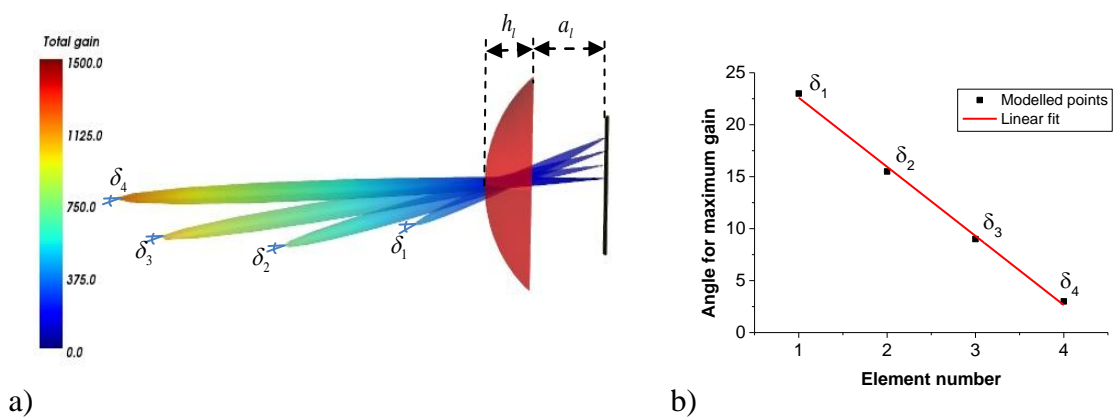

**Fig. 10** a) Modelled system with the lens. Each element was modelled independently and then results were plotted on one figure. The gain is plotted in linear coordinates instead of dB. b) The angle of maximum gain versus the element position for the system with the lens (only the first four elements are shown because of the mirror symmetry). The substrate parameters are $h$=50 μm, $\varepsilon_r$=2, the incident radiation frequency is 300 GHz

First, the lens focus was determined by a direct simulation (the point with maximum intensity is the focus when lens is illuminated by a plane wave). The linear detector array on the substrate with $\varepsilon_r =2$ and $h=50$ μm was placed in the focal plane of the lens, and the simulation for the array together with the lens has been performed to determine the gain diagrams of each antenna of the array. The dependence of the angle of maximal gain on the element number is linear, as **Fig. 10**b demonstrates. The angular position (deviation from the normal) of the major lobe of each FPA element depends linearly on its distance from the optical axis of the lens. Thus, one may expect that the FPA system with the lens does not introduce spatial distortions to the image. However, the lens did introduce distortions in the gain magnitude, which varied from 1300 for central elements of FPA to 400 for those at the edge of the array, while without the lens the gain magnitudes were uniform for the case considered (compare with **Fig. 3** and **Fig. 6**).

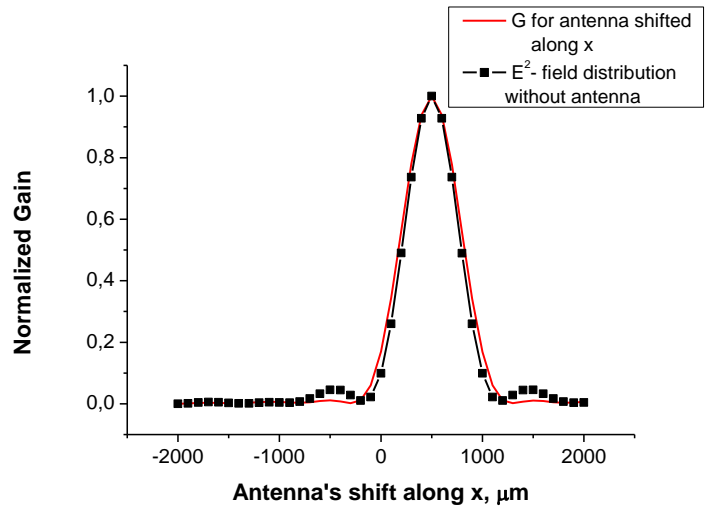

**Fig. 11** The modeled antenna signal compared to the actual intensity distribution. The lens focus is at x=500 μm. The substrate parameters are $h$=50 μm, $\varepsilon_r$ =2

In **Fig. 11** the signal intensity distribution in the focal plane of the lens (the intensity in the Airy disk cut along the *x* direction) is compared to the gain dependence on the antenna shift along the *x* axis obtained for the 4[th] antenna. The antenna was modeled in the transmission mode together with the lens having the same parameters as in **Fig. 10**a. The two distributions practically coincide, so despite the finite dimensions of the antenna, it operates like a point receiver, and correctly reproduces the intensity distribution in the Airy disk. Thus this system can be used for THz imaging in scanning mode. The larger the aperture of the lens, the greater is its gain. Large lens has sharper antenna patterns and could be modeled by methods of quasi-optics [50].

## 6. Conclusions

The substrate electric thickness in THz FPAs plays a crucial role in the frequency characteristics of the system, and should be much lower than the conventional one for CMOS technology. When the substrate is thick and its permittivity is high, a lot of side lobes appear, and the antenna patterns as well as frequency dependencies become different for different FPA elements. The gain dependence on the detector position on the finite substrate is the main factor of optical responsivity $R_V^{opt}$ difference for the detectors in the array, the dependence on the impedance matching coefficient is much weaker. To achieve the uniform multielement system sensitivity, substrates with low $\varepsilon_r$ (<5) and/or low thickness (≤50 μm for Si substrates) should be used. The variation of sensitivity of CMOS FPA elements on relatively thick ($h > 50$ μm) dielectric substrates with high permittivity (e.g., silicon ones) can be explained by the effect of finite size substrate on the antenna gain. Comparison the frequency of the 1st mode of infinite substrate with the frequency dependency of the gain shows that excitation of substrate modes is the reason for the antenna pattern degradation. Antennas on thicker substrate may have better matching with the transistor.

It was demonstrated that the multielement Si CMOS system (with substrate thickness $h = 50$μm and $\varepsilon_r = 2$) can operate as FPA, as well as in devices with mechanical scanning, i.e., that the image is not distorted due to the coupling of the focusing lenses and antennas. However, the incorporation of a lens into simulation introduces distortions in the gain magnitude, which vary from 1300 for central elements of FPA to 400 for those at the edge of the array, while without the lens they are uniform.

The issue of non-homogeneity of detector parameters in multielement FPA of the THz range can be solved in two different ways. The first way is to make each detector with its antenna on its own small-size rectangular dielectric substrate, and to implement FPA by flip-chip technology hybridization on common substrate with low permittivity. The second one is to create an electrically thin substrate using interconnection metallization layers. The latter way is the most appropriate one in terms of the standard CMOS technology.

## 7. Acknowledgments

This work is partly supported by the SPS:NUKR.SFP 984544 Project and a joint grant 01-02-2012 from the National Academy of Sciences of Ukraine and Russian Academy of Sciences.